\documentclass{style/nseJournal}

\usepackage{graphicx}%
\usepackage{multirow}%
\usepackage{amsmath,amssymb,amsfonts}%
\usepackage{amsthm}%
\usepackage{mathrsfs}%
\usepackage[title]{appendix}%
\usepackage{xcolor}%
\usepackage{textcomp}%
\usepackage{manyfoot}%
\usepackage{booktabs}%
\usepackage{algorithm}%
\usepackage{algorithmicx}%
\usepackage{algpseudocode}%
\usepackage{listings}%

\begin{document}

\title{Unstructured Mesh Tools for
Fusion Energy System Design} 

\correspondingEmail{shephard@rpi.edu}
\addAuthor{Mark S. Shephard}{a}
\addAuthor{Jacob S. Merson}{a}
\addAuthor{Onkar Sahni}{a}
\addAuthor{Cameron W. Smith}{b}
\addAuthor{Usman Riaz}{b}
\addAuthor{Fuad Hasan}{a}
\addAuthor{Aditya Y. Joshi}{a}
\addAuthor{Dhyanjyoti D. Nath}{a}
\addAuthor{Abhiyan Paudel}{a}

\addAffiliation{a}{Scientific Computation Research Center, and \\
Department of Mechanical, Aerospace, and Nuclear Engineering\\
Rensselaer Polytechnic Institute,
110 8th St., Troy, NY 12180}
\addAffiliation{b}{Scientific Computation Research Center \\
Rensselaer Polytechnic Institute,
110 8th St., Troy, NY 12180}

\addKeyword{fusion system simulation}
\addKeyword{geometry and meshing}
\addKeyword{code coupling}
\addKeyword{unstructured mesh particle methods}

\titlePage

\begin{abstract}
The execution of accurate simulations of fusion energy systems requires the appropriate representation of critical component geometries as well as the coupling of complex fusion physics codes with one another and with engineering analysis tools. 
This paper examines the challenges of creating simulation workflows that fully leverage existing fusion research codes while integrating them with commercial computer-aided engineering (CAE) software.  Key areas addressed include: (a) the construction and meshing of analysis geometries taking full advantage of available geometric modeling and meshing technologies; (b) the effective coupling of fusion physics and engineering analysis codes; and (c) the support for simulation workflows that couple particle and continuum modeling methods. 

\end{abstract}

\section{Introduction}\label{Sec:Introduction}

The fusion energy industry is dedicating substantial effort to collaborating with the fusion physics research and computer-aided engineering communities to meet system design needs and to develop digital twins of fusion energy systems \cite{schissel2025summary}. 
Although progress has been made, it remains difficult and costly to include full geometric detail of critical components in fusion physics simulations. Another ongoing challenge is the reliable coupling of complex fusion physics codes with one another and with engineering analysis tools used by many fusion companies.

This paper addresses three areas in which the adoption of simulation automation tools can advance the fusion energy industry’s ability to perform accurate, high-fidelity design simulations. 
First, it discusses the construction of CAD-based analysis geometries that incorporate both detailed component geometry and physics-specific geometry, such as magnetic flux surfaces, to create robust analysis-ready CAD models. These models can be readily annotated with mesh control information,
enabling the automatic generation of well controlled meshes for the required analysis codes. Second, considered is given to the coupling of analysis codes into efficient simulation workflows. 
A methodology is outlined for coupling fusion physics codes to one another and to engineering analysis codes, while ensuring accuracy of the transferred fields. Third, it addresses multiscale simulations that combine particle-based and continuum-based methods across complex geometries. In particular, it outlines an infrastructure for particle methods built on unstructured meshes and describes how this approach can be applied to fusion-related analysis codes.

\section{Analysis Geometry and Meshes}\label{Sec:GeometryMeshing}

Most fusion energy system simulations rely on a spatial discretization of the physical domain, referred to as a mesh, over which calculations are performed. Modern computer-aided engineering (CAE) software can automatically generate high-quality unstructured meshes when provided with a properly prepared geometric model. As a result, many newer fusion simulation codes are being designed to operate on unstructured meshes.


Even when suitable simulation codes are available, the process of constructing appropriate meshes remains a significant challenge. This challenge does not stem from mesh generator limitations, but rather from the absence of a proper high-level geometric definition of the analysis domain.
When such a domain definition is available, the full range of simulation automation technologies, including automatic mesh generation, geometry-consistent mesh adaptation and generalized simulation code coupling~\cite{beall2003accessing, shephard2024unstructured, merson2025pcms}, can be effectively applied. The most versatile domain representation capable of supporting the full range of analysis geometry configurations is a non-manifold boundary representation~\cite{weiler1985edge}.  

To support the construction of the analysis geometric models required for fusion system physics and engineering simulations, the analysis geometry often needs to be constructed from geometry provided from multiple sources including:
\begin{itemize}
    \item \textbf{Parameterized CAD models} that support design optimization. Examples include blankets systems defined with ParaStell~\cite{moreno2024parastell} and magnetic coils represented as a wire framework~\cite{hammond2025framework}, as well as developing RF antenna models for heating in magnetically confined systems and divertor system models for heat extraction and extraction of impurities.
    \item \textbf{Manufacturing CAD models}, which contain detailed representations of all components. These models often contain geometric features that may be unnecessary for a given simulation and can therefore be removed, such as bolts in antenna assemblies or sliver faces introduced during CAD healing operations~\cite{shephard2024unstructured}.
    \item \textbf{Constructive solid geometry models} such as those commonly in neutronics calculations.
    \item \textbf{Physics-defined components}, such as flux curves and/or surfaces, X-points and O-points derived from equilibrium field solutions using tools such as EFIT~\cite{lao1985reconstruction} for tokamaks, and VMEC \cite{hirshman1983steepest} and DESC~\cite{dudt2020desc, dudt2022webpage} for stellarators. 
    \item \textbf{Segmented image data} used support analyses of ``as-built'' or ``as-is'' geometry at specific stages in system life cycle. 
    \item \textbf{Already discretized geometry} including, faceted surfaces (e.g., STL files) or surface triangulations derived from previously defined meshes.  
\end{itemize}

Within the research environments where many complex fusion physics codes have been developed, the analysis geometric definition has traditionally been tailored to match the immediate needs of the simulation codes and their associated meshing tools. For example, the highest-level model understanding given in many codes is a set of mapped rectangles as used to support a block structured finite difference calculation. While expedient, this strategy typically limits the level of geometric complexity that can be supported and does not effectively enable the execution of coupled multi-fidelity analyses required during the design of fusion energy systems. 

In contrast, fusion energy companies, whose ultimate objective is the design and construction of deployable fusion systems, generally prefer to employ a common, high-level geometry definition. 
Such a definition supports the execution of design simulations as well as the development of fusion system digital model(s). 

The boundary representation kernels underlying CAD systems used in manufacturing industries can support the requirements of analysis domain geometries. By leveraging existing CAD and computer-aided engineering (CAE) geometry interaction and interoperability tools, the full range of geometry requirements encountered in fusion system simulations can be supported.

Parameterized geometric models define key geometric components in terms of parameters that that can be varied within specific limits, with models automatically updated by design optimization procedures.  
Example design optimization procedures used in stellarator design are STELLOPT~\cite{lazerson2013stellopt, lazerson2020stellopt} and SIMSOPT~\cite{landreman2021simsopt} for defining optimal plasma shapes, and coil optimization~\cite{hammond2025framework}.
While the geometric shape information associated with current parametric models can be complex, the topology remains relatively simple. However, when considering other components such as blankets, RF antennas and island divertors, the level of topological complexity increases substantially and remains an area that is not yet well addressed. For example, the parameterizing the outer surface of a blanket is easily supported; however, incorporating the numerous ports that penetrate the blanket introduces significant additional complexity. Similarly, the parametric definition of an island divertor is reasonably straight forward if the objective is limited to targeting a uniform surface temperature, but the complexity of both the parametric model and the associated optimization increases markedly when additional requirements such as particle exhaust and particle retention are considered. Further development efforts that exploit the full capabilities of CAD systems to define parametric models accounting for the complete range of component functions and constraints are needed.

\begin{figure}
\centering
\captionsetup{width=\linewidth, justification=centering}
    \includegraphics[width=0.8\linewidth]{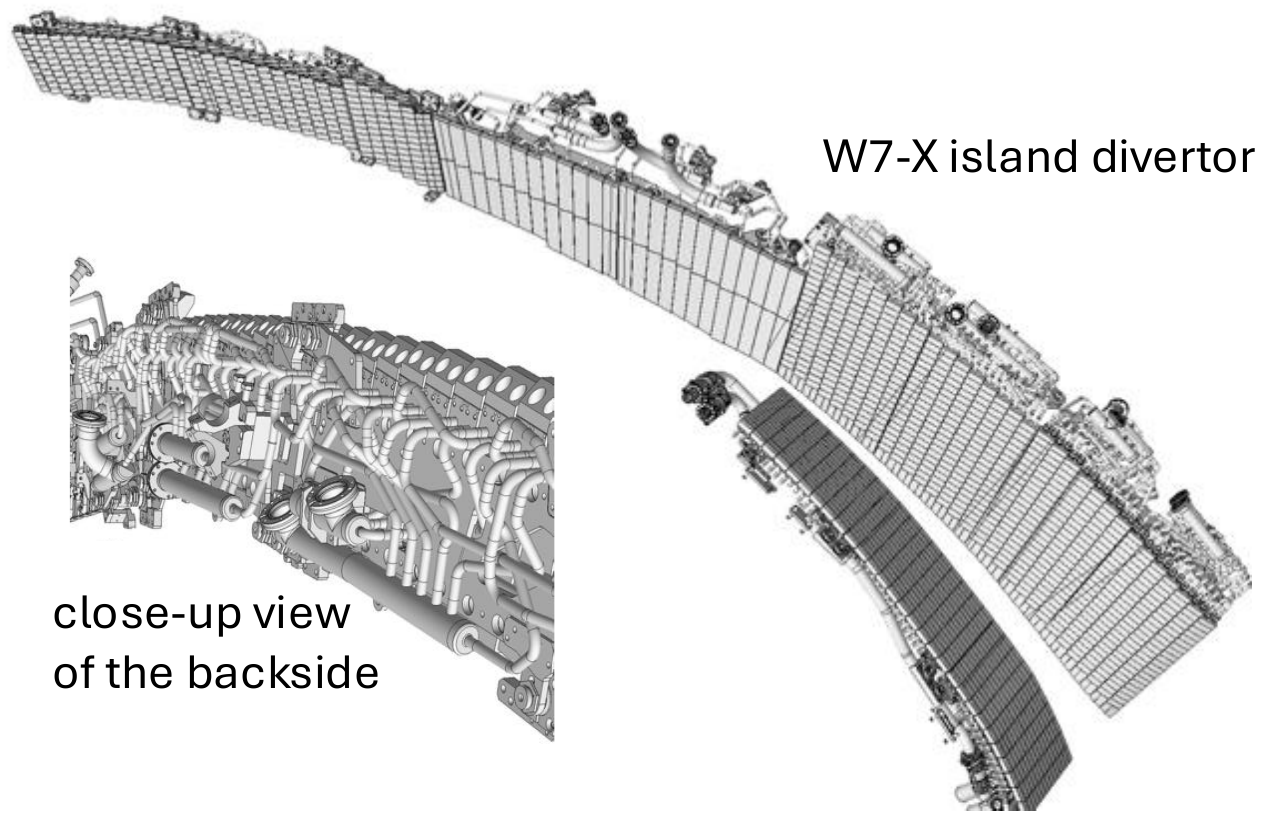}
    \caption{CAD model of the W7-X island divertor generated from Max-Planck-Institute for Plasma Physics, 2024, CAD print files of Wendelstein 7-X (https://www.ipp.mpg.de/5440442/cad\_druck, accessed on 27 January 2026).}
    \label{fig:divertorCAD}
\end{figure}

\begin{figure}
\centering
\captionsetup{width=\linewidth, justification=centering}
    \includegraphics[width=0.8\linewidth]{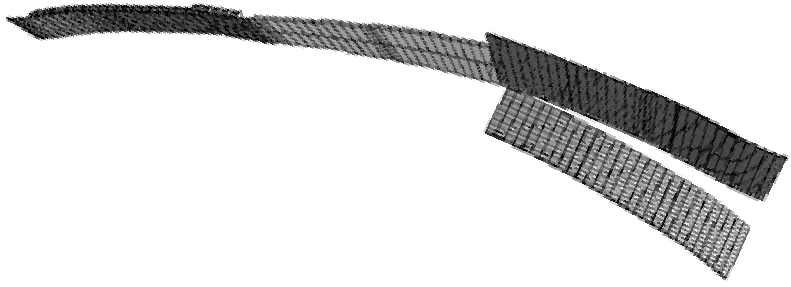}
    \caption{Isolated island divertor.}
    \label{fig:single-divertorCAD}
\end{figure}

\begin{figure}
\centering
\captionsetup{width=\linewidth, justification=centering}
    \includegraphics[width=1.\linewidth]{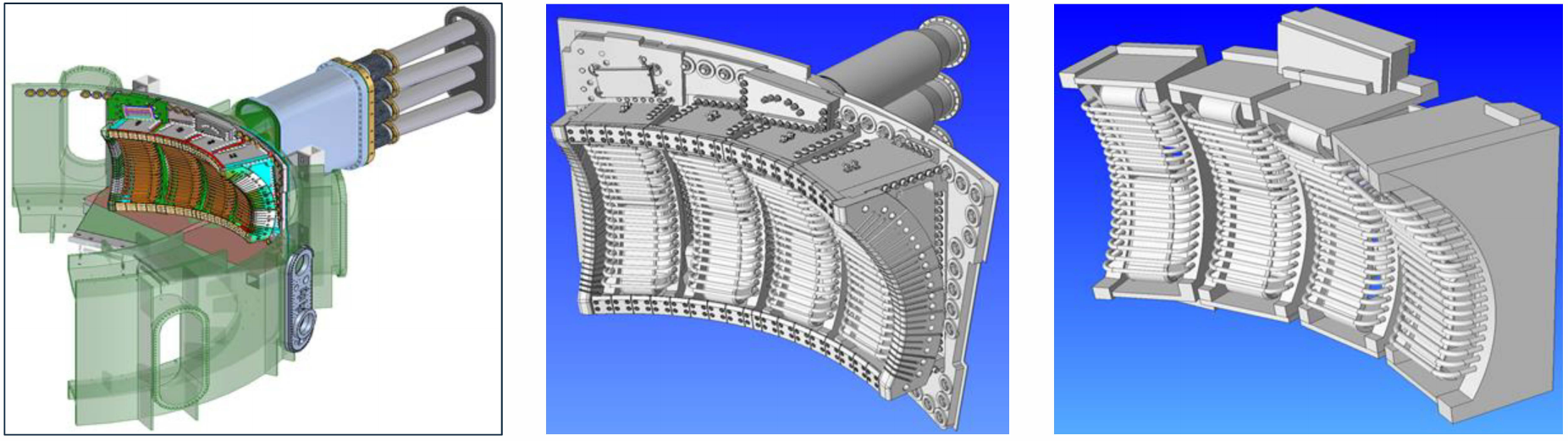}
    \caption{Manufacturing CAD (left), extracted components (middle), simplified Analysis Geometry (right).}
    \label{fig:CAD-to-analysis-geo}
\end{figure}

 In the case of fusion system components that have already been designed, all the geometry of potential interest will be included in the manufacturing CAD model. One challenge associated with these models is that they often contain many parts that are not relevant to a given simulation. For example, consider the analysis of heat loads on the W7-X island divertor using the divertor CAD model available from the Max Planck Institute for Plasma Physics~\cite{W7-XCAD}. 
 Figure~\ref{fig:divertorCAD} shows the island divertor CAD model. This model includes thousands of parts, whereas for the purpose of a heat load analysis only the geometry shown in Figure~\ref{fig:single-divertorCAD} is required. Even when the components of interest have been extracted from the manufacturing CAD, additional model modifications that are needed. The most obvious modifications are simplifications to eliminate geometric model pieces that are not relevant to the analysis. Including such features in the analysis model would significantly increase the number of elements generated and, consequently, the overall simulation cost. Figure~\ref{fig:CAD-to-analysis-geo} illustrates an example of an antenna component used in RF and impurity transport analyses. The left image shows the supplied manufacturing CAD of the antenna assembly, the middle image shows the extracted components of interest, and the right image shows the simplified geometry used for analysis.

\begin{figure}
\centering
\captionsetup{width=\linewidth, justification=centering}
    \includegraphics[width=1.0\linewidth]{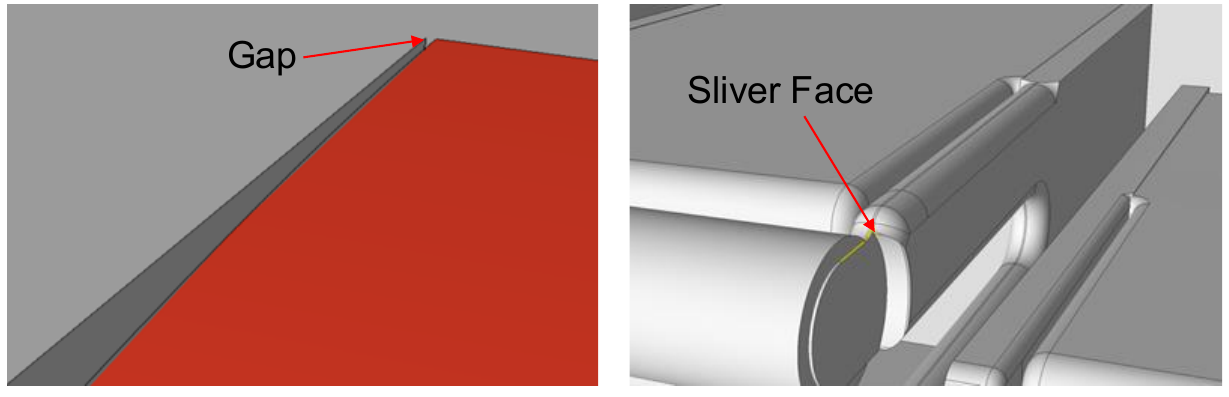}
    \caption{Problematic model assembly features.}
    \label{fig:Slivers-Gaps}
\end{figure}

A key step in going from the middle image in Figure~\ref{fig:CAD-to-analysis-geo} to the right image is model cleanup to address small geometric imperfections of three types: overlaps, gaps, and sliver faces (see Figure~\ref{fig:Slivers-Gaps}). Overlaps and gaps arise during the assembly of components due to imprecise positioning and/or differences in component tolerances. Sliver faces are typically introduced by CAD systems when gaps between components are detected.
Although some CAE systems~\cite{shephard2024unstructured, simmetrix_web} can automatically eliminate up to 95\% of these model imperfections, experience~\cite{shephard2024unstructured} indicates that the interactive removal of the remaining 5\% is time-consuming, as the associated features are extremely small and difficult to locate. There are ongoing developments of a new approach that is expected to enable the robust elimination of 100\% of overlaps, gaps, and sliver faces.

The second step in converting the cleaned-up model into the desired analysis geometry is defeaturing, which refers to the process of eliminating geometric features that are irrelevant to the calculation of quantities of interest in a given analysis. The ability to automatically or semi-automatically remove such undesired features is greatly enhanced when feature-based models are available. Feature-based models include functional identifiers associated with geometric features, facilitating their identification and removal.
 
\begin{figure}
\centering
\captionsetup{width=\linewidth, justification=centering}
    \includegraphics[width=1.0\linewidth]{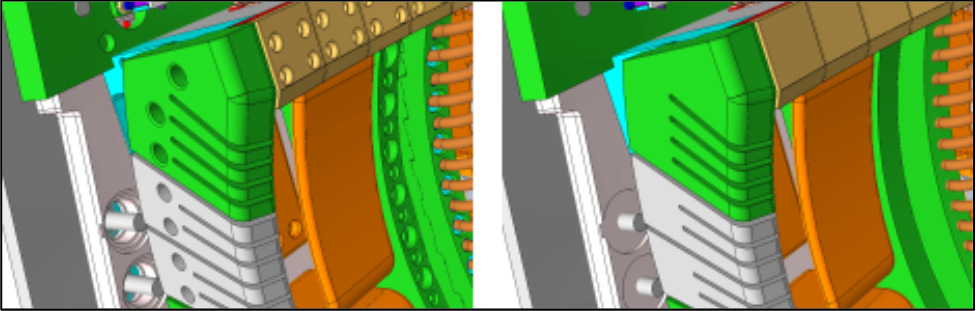}
    \caption{Defeaturing unneeded geometric model features.}
    \label{fig:Defeaturing}
\end{figure}

 In cases where feature identification information is not available, geometric reasoning can be applied to help identify and eliminate candidate features. Care must be taken, as feature size alone is not a sufficient criterion. For example, consider the close-up of a portion of an RF antenna shown in the left image of Figure~\ref{fig:Defeaturing}, where small geometric features include bolt holes and slots in the limiter. The bolt holes are not relevant to the RF simulations; however, the limiter slots play a critical role in the definition of the RF fields and must be retained in the analysis geometry.
To accelerate the removal of undesired geometric features, interactive tools have been developed~\cite{simmetrix_web, tendulkar2018unstructured} that allow the user to select a candidate feature for elimination. Geometric reasoning is then applied to identify and remove all equivalent features. The right image of Figure~\ref{fig:Defeaturing} shows a close-up of the RF antenna with all bolt holes eliminated after the interactive selection of one hole of each size.  

\begin{figure}
\centering
\captionsetup{width=\linewidth, justification=centering}
    \includegraphics[width=1.0\linewidth]{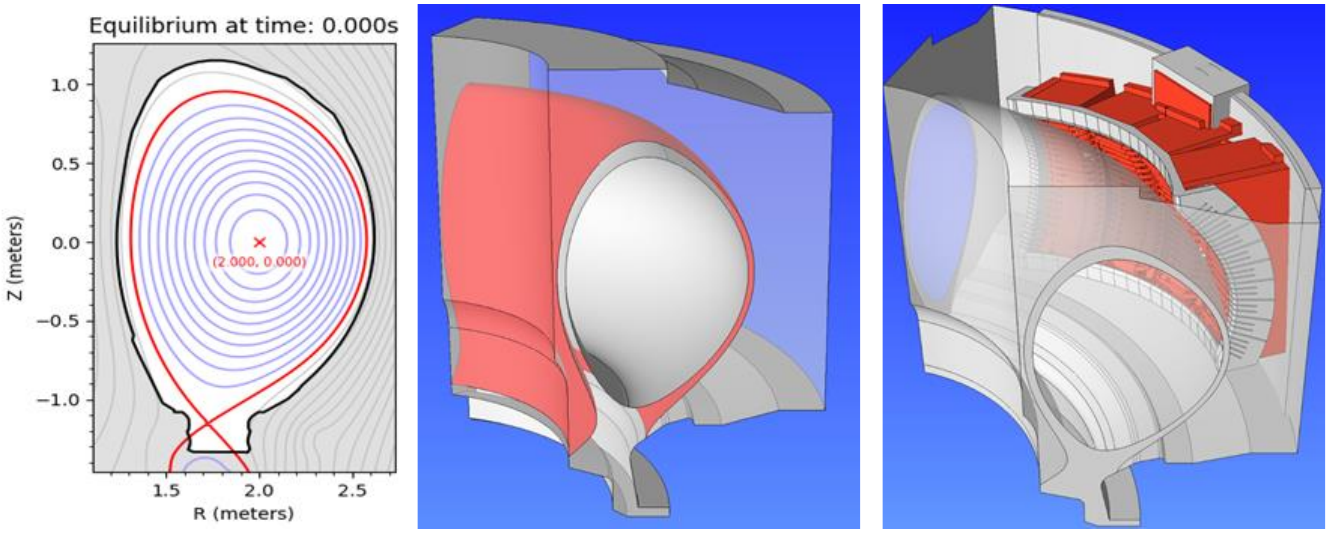}
    \caption{Assembling components into the analysis model.}
    \label{fig:Assemble-Model}
\end{figure}

The final step in constructing an analysis CAD geometry is the assembly of component geometries. In fusion system physics analyses, these components often originate from multiple sources. Figure~\ref{fig:Assemble-Model} shows a representative example in which the analysis CAD model is assembled from three distinct geometry inputs: physics geometry defined by magnetic field data, discretized geometry representing the tokamak wall, and an antenna analysis CAD model derived from the manufacturing CAD of the antenna assembly.
In this example, the magnetic field data are extracted from a G-EQDSK~\cite{G-EQDSK} file containing EFIT equilibrium data, provided as values on a background grid. The magnetic field over this grid is represented using cubic splines. Selected flux curves are constructed as cubic spline contours of the flux function. The flux surfaces, shown in the center and right images of Figure~\ref{fig:Assemble-Model}, are defined by extruding these contours in the toroidal direction~\cite{riaz2024modeling, zhang2023development}. In the case of stellarator flux surfaces, advantage can be taken of the ability of CAD systems to incorporate external geometric shape information. In this case, flux surface geometry can be defined directly using the Fourier representation extracted from VMEC equilibrium files.
The tokamak wall geometry shown in Figure~\ref{fig:Assemble-Model} is constructed from a piecewise linear discretization by applying geometric reasoning to identify wall segments that should be concatenated to form linear edges and those that should be combined and represented as curved edges. This process results in a geometrically consistent set of edges defining a closed wall loop composed of straight edges and curved edges, with the latter defined by cubic spline fits to the nodes of the selected segments~\cite{zhang2023development}. The wall curve loop is then extruded in the toroidal direction.
Finally, the flux and wall surfaces are combined with the antenna component, and the additional volume regions to be meshed are identified, resulting in the final analysis CAD model shown in the right image of Figure~\ref{fig:Assemble-Model}.

Employing a boundary representation for both the analysis geometric model and the mesh supports the effective application of mesh adaptation through local mesh modification~\cite{li20053d}. A boundary representation consists of two components. The first is a graph in which the graph nodes represent topological entities and the graph edges represent selected adjacencies of the entities. An example upward adjacency is the one or two regions that a given face bound, while a common downward adjacency is the set of edges bounding a given face. The second component is the geometric shape information that is associated with appropriate topological entities in the boundary representation.

Analysis CAD models can contain regions with multiple shells and faces with multiple loops. To account for this, geometric model boundary representations include regions, shells, faces, loops, edges, and vertices. Since analysis geometric models may consist of general combinations of different material regions and reduced-dimensional entities (e.g., dangling faces that do not bound any region), a general non-manifold topological representation is required~\cite{weiler1985edge, Weiler1986}.
In this work, it is assumed that the analysis model is defined within geometric modeling software such as Parasolid~\cite{Parasolid}, ACIS~\cite{ACIS}, or OpenCascade~\cite{Open_CASCADE}. In these systems, the shape information associated with model faces and edges is most commonly represented using non-uniform rational B-splines defined in a local parametric coordinate system and mapped to a global Cartesian coordinate system.
However, other shape representations are also possible. For example, in one ongoing fusion system modeling project, the surface geometry is defined in terms of Fourier coefficients derived from VMEC files. Since geometric modeling systems support Boolean operations on model components, surfaces may be trimmed, and regions generally cannot be described using a single simple parametric coordinate system.

Geometric modeling kernels provide application programming interface (API) operations that allow programmers to perform specific geometric queries. The mesh adaptation procedures presented in this paper use the geometric modeling system API to interact with the shape information of the analysis geometric model~\cite{beall2003accessing}. The API functions support a range of topological queries (such as determining which material regions lie on either side of a surface) and geometric queries (such as computing the normal to a surface or identifying the material region that contains a given point).

\begin{figure}
\centering
\captionsetup{width=\linewidth, justification=centering}
    \includegraphics[width=.4\linewidth]{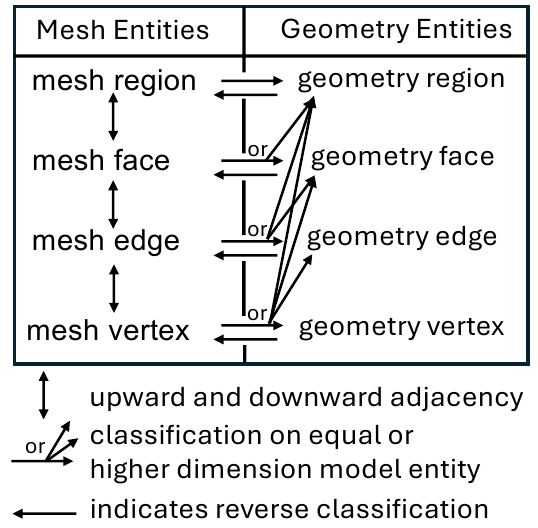}
    \caption{Mesh topology and relationship of geometric model and mesh entities.}
    \label{fig:classification}
\end{figure}

For the mesh topology, it is assumed that each mesh region (3D volume element) is bounded by a single shell of adjacent mesh faces, and that each mesh face is bounded by a single loop of adjacent mesh edges. Under this assumption, a complete mesh boundary representation need only include region, face, edge, and vertex entities, along with a set of mesh adjacencies that is complete in the sense that any of the 12 possible first-order adjacency operations can be evaluated in $O(1)$ time.
Reference~\cite{beall1997general} discusses options for complete mesh topological representations. The left portion of Figure~\ref{fig:classification} illustrates the mesh entities and the use of one-level mesh adjacencies~\cite{beall1997general}, indicated by vertical double arrows, which together form a complete representation.

A key requirement for enabling a broad range of simulation automation technologies is the maintenance of appropriate relationships between mesh entities and analysis model topological entities. In particular, supporting automatic mesh generation and mesh adaptation requires preserving the association between mesh topological entities and the primary topological entities of the analysis geometric model. One effective method for achieving this, referred to as \textit{mesh entity classification}~\cite{beall2003accessing, beall1997general}, assigns to each mesh entity knowledge of the highest-order analysis model topological entity within which it lies or of which it is a part.
As indicated by the right-pointing and upward-angled arrows in Figure~\ref{fig:classification}, classification is defined as follows. A mesh region is classified as lying within a geometric model region. A mesh face is classified as lying either within a geometric model region or on a geometric model face. A mesh edge is classified as lying within a geometric model region, on a geometric model face, or on a geometric model edge. Finally, a mesh vertex is classified as lying within a geometric model region, on a geometric model face, on a geometric model edge, or at a geometric model vertex.

In addition to mesh entity classification, another useful relationship between geometric model topological entities and mesh topological entities is \textit{reverse classification}. Reverse classification identifies, for a given model entity, the set of mesh entities of equal order that are classified on it. This relationship is valuable for tasks such as solution setup and other related operations. The left-pointing arrows in Figure~\ref{fig:classification} indicate reverse classification.

Given an analysis CAD model in which the model entities are attributed with mesh control information, controlled unstructured meshes can be automatically generated using available mesh generators~\cite{simmetrix_web, geuzaine2009gmsh, gmsh_web}. Beyond initial mesh generation, mesh adaptation tools can be coupled to the analysis codes to adapt the mesh during the simulation to ensure control of the solution accuracy~\cite{shephard2024unstructured} and/or account for evolving geometry~\cite{yang2022parallel}. 

\begin{figure}
\centering
\captionsetup{width=\linewidth, justification=centering}
    \includegraphics[width=0.5\linewidth]{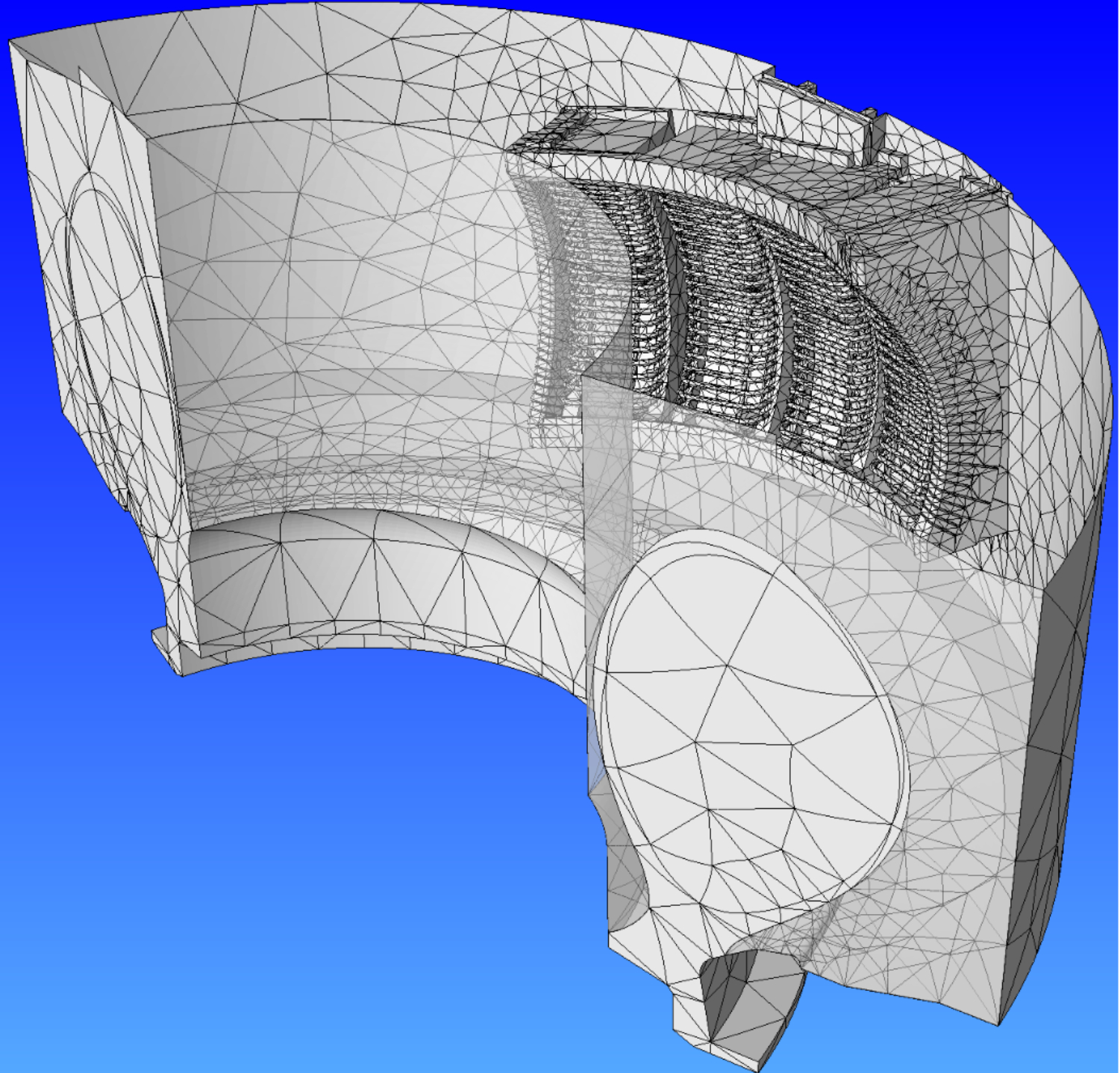}
    \caption{Curved mesh suitable for high-order finite element RF analysis.}
    \label{fig:PetraM-mesh}
\end{figure}

Figure~\ref{fig:PetraM-mesh} shows a coarse, curved mesh generated for an RF simulation of the analysis CAD model shown in the right image of figure~\ref{fig:Assemble-Model}. Coarse, curved meshes are preferred for these simulations because they facilitate the use of high-order finite element methods~\cite{kolev2021efficient}, which are optimal for high-performance RF analyses~\cite{siboni2022adaptive}. 

\begin{figure}
\centering
\captionsetup{width=\linewidth, justification=centering}
    \includegraphics[width=0.8\linewidth]{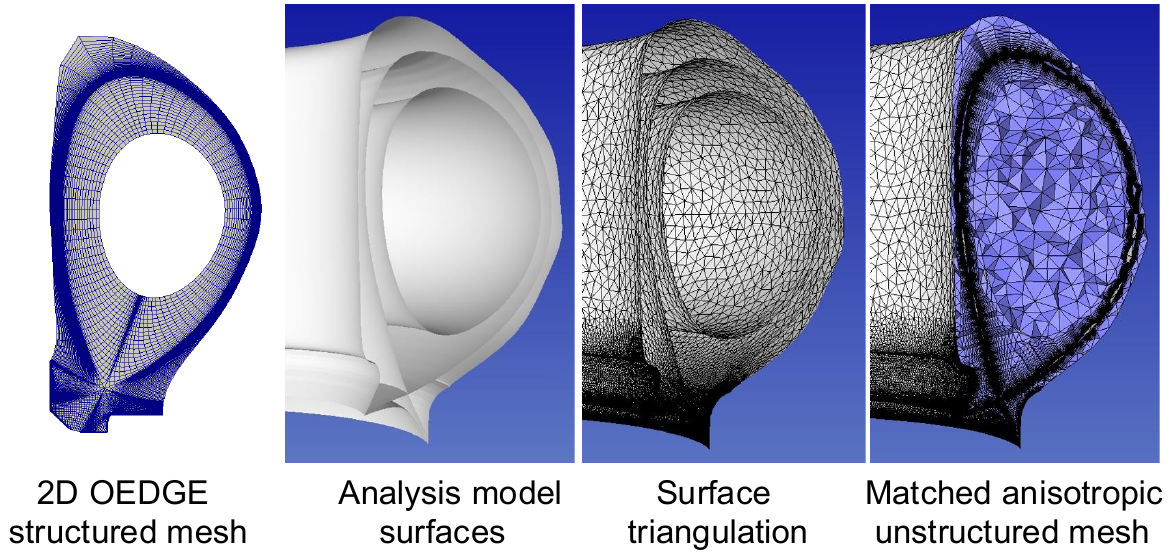}
    \caption{Anisotropic unstructured mesh sized for accurate transfer of background fields from 2D OEDGE structured mesh for GITRm impurity transport analysis.}
    \label{fig:GITRm-anisotropic-mesh}
\end{figure}

\begin{figure}
\centering
\captionsetup{width=\linewidth, justification=centering}
    \includegraphics[width=0.6\linewidth]{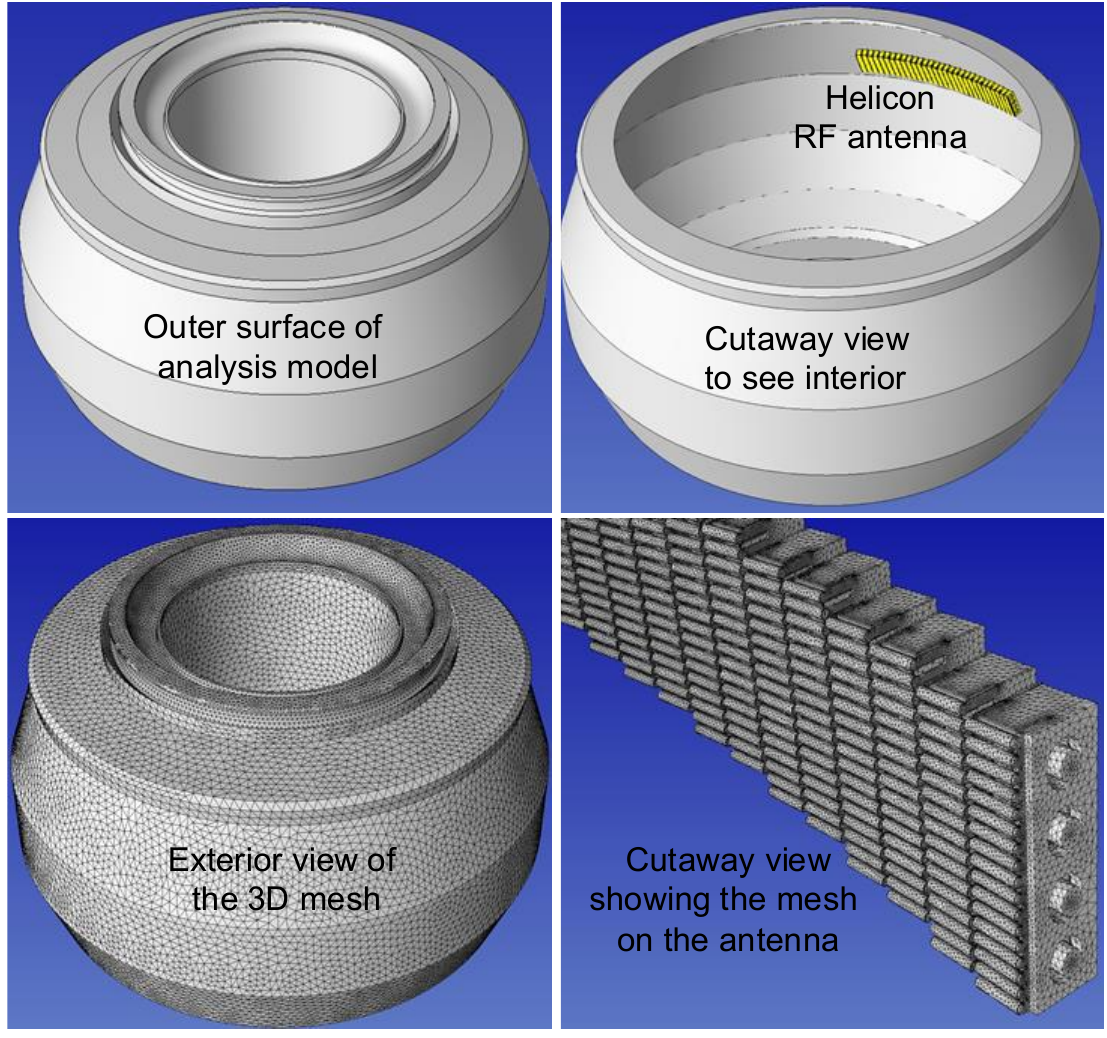}
    \caption{GITRm 3D mesh for impurity transport simulation of DIII-D with Helicon RF antenna included.}
    \label{fig:GITRm-antenna}
\end{figure}

For impurity transport simulations, efficient and accurate solution field transfer requires anisotropically graded meshes that match the anisotropic meshes used to calculate the background fields governing impurity motion. Figure~\ref{fig:GITRm-anisotropic-mesh} demonstrates the generation of such meshes for impurity transport simulations using GITRm~\cite{nath20233d, nath2025gpu}.
The far-left image shows the 2D OEDGE mesh used to define the axisymmetric background field. The next two images show the surfaces on which mesh control information is applied and the resulting finite element mesh. To match the anisotropic nature of the OEDGE mesh, mesh control information is specified to generate a semi-structured anisotropic mesh extending into the regions on both sides of the surface. The right image shows the automatically generated unstructured mesh based on this mesh control information.
Since GITRm employs a fully 3D mesh, analyses that account for fully three-dimensional behavior are possible. Figure~\ref{fig:GITRm-antenna} shows views of the geometry and mesh generated for GITRm analyses of DIII-D with a Helicon RF antenna~\cite{kumar2026understanding}.

\section{Code Coupling and the Transfer of Solution Fields}\label{Sec:SolutionTransfer}
The creation of fusion energy system simulation workflows often requires coupling solution fields between complex physics codes, each of which may employ distinct parallel data structures, discretization schemes, and numerical methods. Experience gained in the Exascale Computing Project demonstrated that field transfer between codes must be performed in a manner consistent with the discretizations used in each code. Moreover, coupling approaches that require modifying one or both codes make it difficult to keep development branches in lockstep with the main branches of the individual codes.

To address the challenge of coupling complex physics codes and ingesting data into AI systems, the Parallel Coupler for Multimodel Simulations (PCMS)~\cite{merson2025pcms} is being developed. Key aspects of PCMS that are critical to meeting the desired simulation requirements include:
\begin{itemize}
    \item Integration of coordinate and data transformations across multidimensional domains to support unique and nonlinear coordinate systems and field data.
    \item Distributed control and transfer algorithms that account for the details of field representations by enabling the use or implementation of native interrogation methods wherever possible.
    \item Non-intrusive operation that makes use of existing data structures and methods.
    \item High-level, discretization-independent specification of simulations supporting general geometric domains and the application of simulation automation tools (see Section~\ref{Sec:GeometryMeshing}).
    \item Flexible execution modes, ranging from file-based one-way coupling to in-memory coupling of massively parallel physics codes, each with their own parallel distribution of data and operations.
\end{itemize}

\begin{figure}
    \centering
    \includegraphics[width=1.0\textwidth]{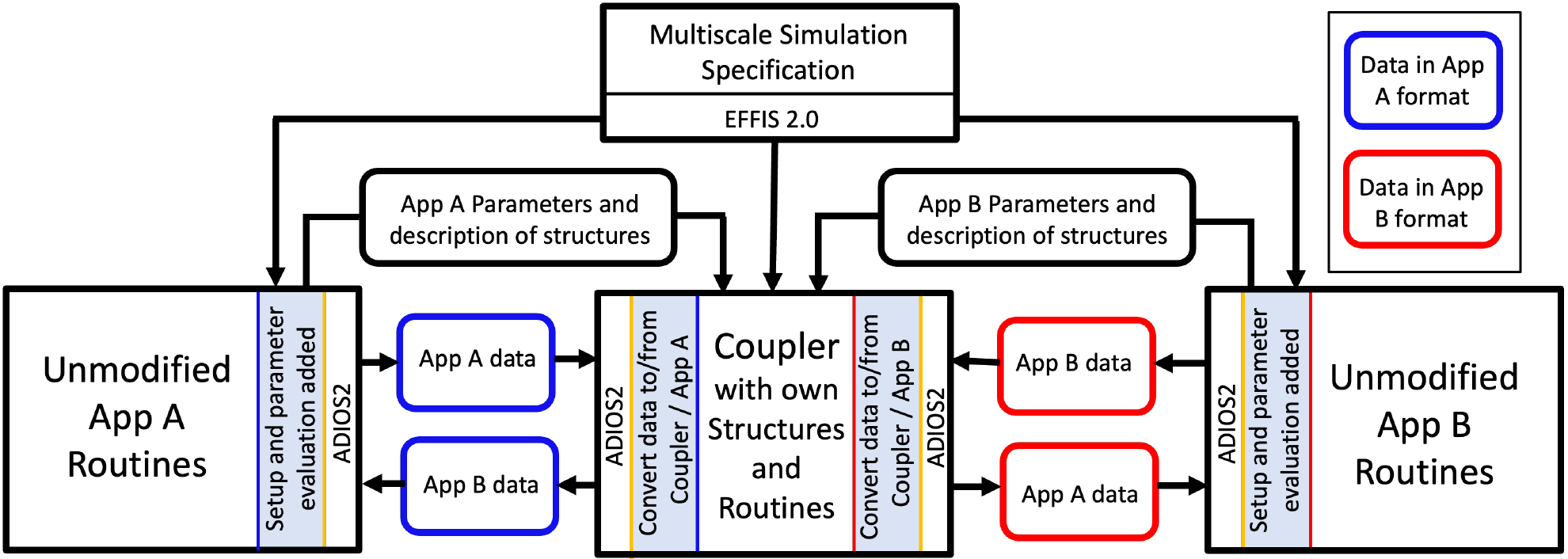}
    \caption{Overview of the PCMS components and operations. Each application uses native data structures and the coupler uses an intermediate representation. Inter-application communication is performed with ADIOS2 and EFFIS provides job control.}
    \label{fig:PCMS-components}
\end{figure}

Figure \ref{fig:PCMS-components} illustrates the component interactions within PCMS. The applications perform their operations on unmodified data structures. Data is prepared for transfer using an adapter class. All data and metadata are sent over the network using ADIOS2~\cite{godoyADIOSAdaptableInput2020}, which supports both in-memory and file-based transport.

Multiple methods exist for transferring solution fields from a sending mesh to a receiving mesh. The accuracy, stability, and physical validity of a transferred solution are strongly dependent on the transfer method employed. The solution transfer methods supported in PCMS can be categorized as \textit{mesh-based methods} or \textit{pointwise methods}. Mesh-based methods use the underlying structure of the discretizations of both the sending and receiving meshes. They are the most complex methods to implement, particularly when either the source or target mesh employs a solution field discretization that has not already been implemented in PCMS.

The simplest approach to implement is interpolation, which assigns to each target mesh point the value of the source field evaluated at the target nodal locations. General interpolation of the source field in a manner consistent with the underlying shape functions requires either the ability to evaluate the application field within the coupler or the ability to perform queries over the network to the source application. Direct interpolation can only enforce conservation constraints through global scaling, which does not account for conservation at any scale other than the entire domain.

When the source field cannot be queried at arbitrary points, but source values are known at specific locations (e.g., a point cloud of nodal locations), two common approaches are used. The first is $0^{\text{th}}$-order interpolation, such as \emph{nearest neighbor}, in which the closest point in the source point cloud is used to assign the target value. This method provides no mechanism to control accuracy or enforce conservation, except through global scaling.

Another common pointwise method that works with point cloud data is the use of shape functions that are independent of the underlying discretization, e.g., radial basis functions (RBFs), along with weighted polynomial fitting~\cite{slatteryMeshfreeDataTransfer2016}. This approach is compelling because it can provide accurate field transfer in a black-box setting and can be efficiently implemented on data-parallel hardware such as GPUs when local basis functions are used. However, it has two practical disadvantages. First, it involves a number of adjustable parameters, such as basis function type, cutoff radius, polynomial order, and regularization, which typically require problem-dependent tuning to achieve satisfactory results. Second, it does not provide a mechanism to control conservation beyond global scaling, since it does not maintain a global partitioning of space. A wide range of radial basis functions has been implemented in PCMS.

Historically, the gold standard among mesh-based methods is \textit{mesh intersection methods}~\cite{jiaoCommonrefinementbasedDataTransfer2004}, which are well suited for satisfying conservation requirements. PCMS supports the application of mesh-intersection-based solution transfer. 
Recently, we have developed a new conservative coupling strategy for black-box coupling in cases where point evaluations of the source field are available, but full discretization information is not available~\cite{merson2025stochastic}. In this method, Monte Carlo integration is used to compute the load vector in a Galerkin projection over the target elements. With a sufficient number of sample points per element, this method can achieve levels of accuracy and conservation comparable to those of mesh intersection methods.

\begin{figure}
    \centering
    \begin{subfigure}{0.45\linewidth}
    \centering
    \includegraphics[width=\linewidth]{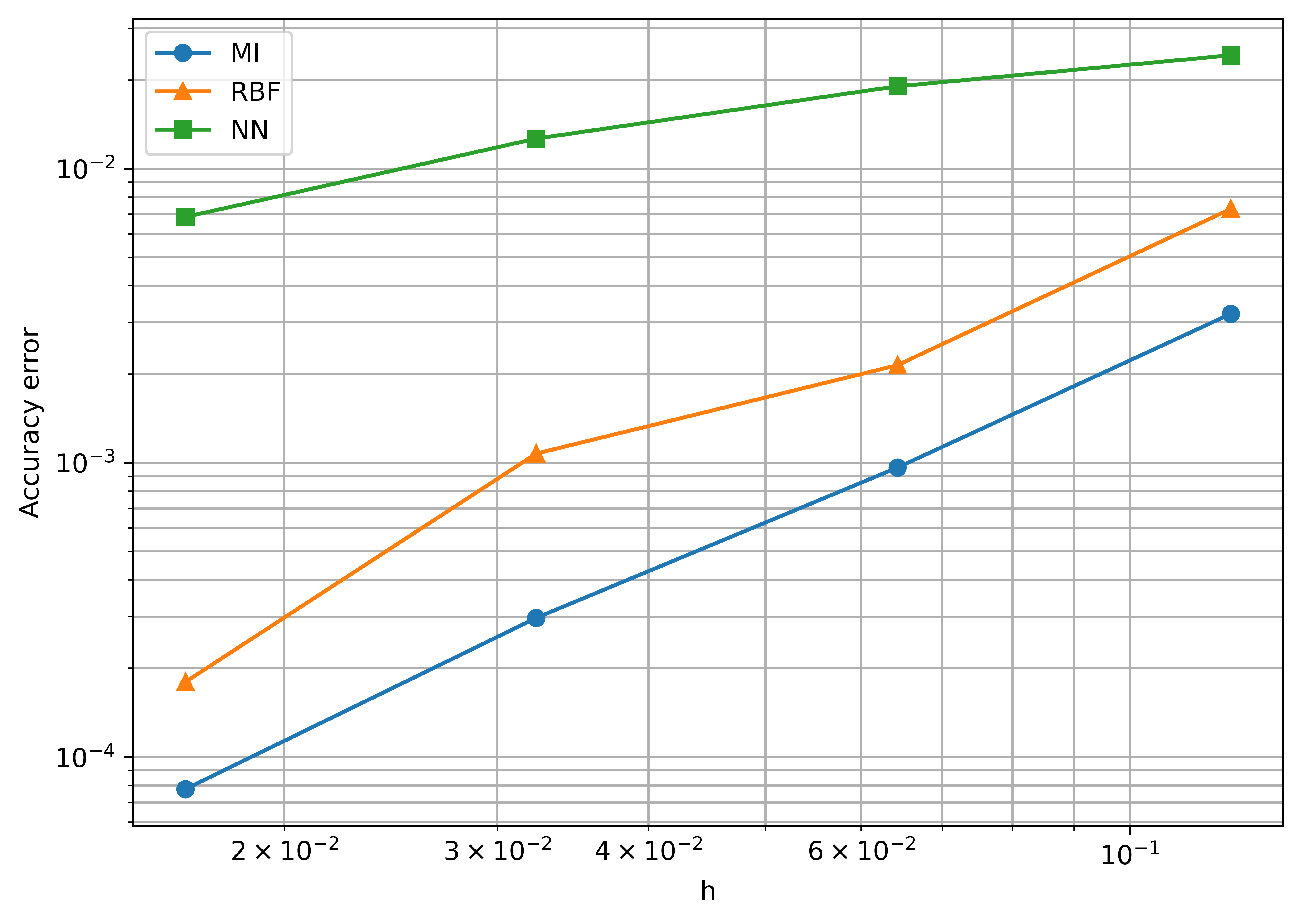} 
    \caption{Accuracy Error}
    \end{subfigure}
    \begin{subfigure}{0.45\linewidth}
    \centering
    \includegraphics[width=\linewidth]{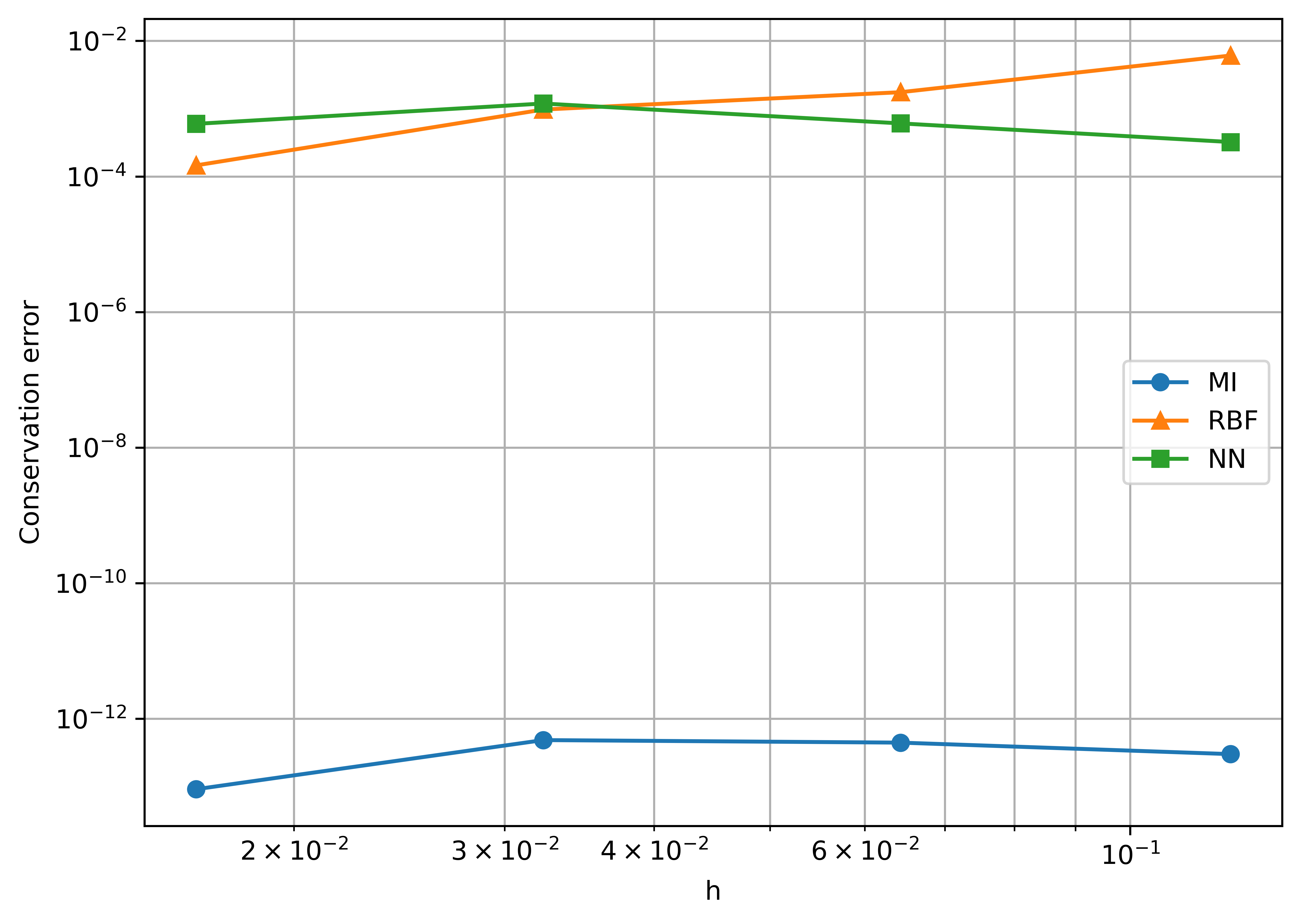} 
    \caption{Conservation Error}
    \end{subfigure}
    \caption{Accuracy and conservation error for the polynomial function given by Eq.~\eqref{eq:polynomial-field} on a square domain with varying element sizes for field transfer performed with mesh intersection (MI), radial basis functions (RBF), and nearest neighbor interpolation (NN).}
    \label{fig:nearest-neighbor-accuracy-conservation}
\end{figure}

Following the approach in \cite{merson2025stochastic} we perform a convergence study showing the accuracy and conservation of the nearest neighbor method, the mesh intersection method, and the radial basis function method for a sequence of target and source mesh pairs with a characteristic mesh size $h$. The meshes are generated by gmsh \cite{geuzaine2009gmsh}. In each case we use linear basis functions and evaluate the polynomial
\begin{equation}
    f(x)=5y^3+x^2+2y+3
    \label{eq:polynomial-field}
\end{equation}
on the source mesh prior to performing the field transfer operation. For the radial basis function we use a $C^4$ basis function \cite{merson2025pcms}, linear local polynomials, and the cutoff radius is adapted to achieve ensure at least 12 source points are within the cutoff radius for each target point.

We evaluate errors on the intersection of the source and target meshes (supermesh) because both the source and target finite element spaces are exactly representable in the supermesh. The accuracy error in the supermesh is given by 
\begin{equation}
E_{L^2}
  =
  \frac{
    \| f^s - f^t \|_{L^2(\Omega)}
  }{
    \| f^s \|_{L^2(\Omega)}
  }
  =
  \frac{
    \left(
      \int_{\Omega}
        \bigl( f^s(x) - f^t(x) \bigr)^2
      \,\mathrm{d}\Omega
    \right)^{1/2}
  }{
    \left(
      \int_{\Omega}
        \bigl( f^s(x) \bigr)^2
      \,\mathrm{d}\Omega
    \right)^{1/2}
  }
\label{eq:relative_L2_error}
\end{equation}
where \(f^s\) is the source field, \(f^t\) is field on the target mesh after field transfer, and \(\Omega\) is the domain.

We also define the conservation error on the supermesh as
\begin{equation}
E_{\mathrm{mass}}^{\mathrm{SM}}
  =
  \frac{
    \left|
      \int_{\Omega} f^s(x)\,\mathrm{d}\Omega
      -
      \int_{\Omega} f^t(x)\,\mathrm{d}\Omega
    \right|
  }{
    \left|
      \int_{\Omega} f^s(x)\,\mathrm{d}\Omega
    \right|
  }.
\end{equation}

Fig.~\ref{fig:nearest-neighbor-accuracy-conservation}a shows the accuracy errors for nearest neighbor, mesh intersection, and radial basis function methods. For the field under consideration, mesh intersection was the most accurate. Although the magnitude of the error was slightly worse with radial basis functions, it maintained similar scaling with the mesh size. Nearest neighbor was almost two orders of magnitude worse than mesh intersection at the finest mesh sizes and did not scale as well as the other methods when the number of elements was reduced.

The conservation error is shown in Fig.~\ref{fig:nearest-neighbor-accuracy-conservation}b. For the field under consideration, the conservation error from the mesh intersection method is eight orders of magnitude less than for the radial basis funcion  and nearest neighbor methods. This clearly demonstrates the value of incorporating control of the conservation errors in the field transfer method. For mesh intersection and nearest neighbor, the conservation error is relatively flat with the mesh resolution. The radial basis function method shows a downward trend in the conservation error with the mesh resolution.

\begin{figure}
    \centering
    \begin{subfigure}{0.4\linewidth}
        \centering
        \includegraphics[width=\linewidth]{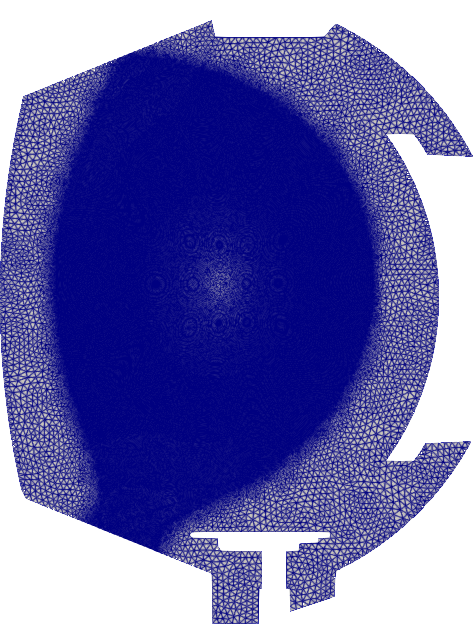}
        \caption{Source mesh}
    \end{subfigure} \hfill
    \begin{subfigure}{0.4\linewidth}
        \centering
        \includegraphics[width=\linewidth]{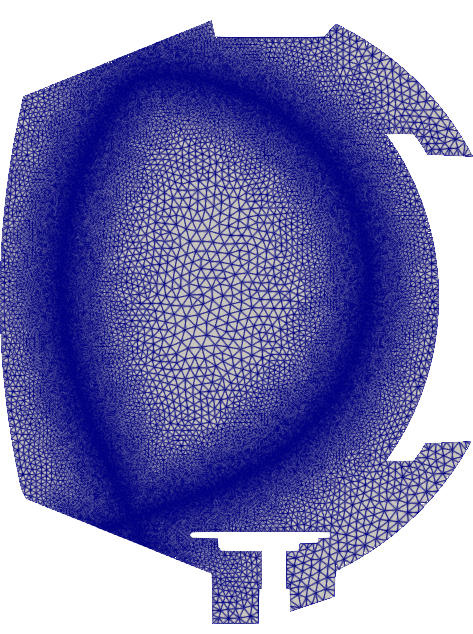}
        \caption{Target mesh}
    \end{subfigure}
    \caption{Source and target meshes on WEST reactor geometry used for field transfer. In the core region of the source mesh (a) is aligned to the magnetic field and has only one element between each flux curve as required for XGCm simulations. The target mesh (b) is a general unstructured mesh over the full domain.}
    \label{fig:west-meshes}
\end{figure}

\begin{figure}
    \centering
    \begin{subfigure}{0.23\linewidth}
    \centering
    \includegraphics[width=\linewidth]{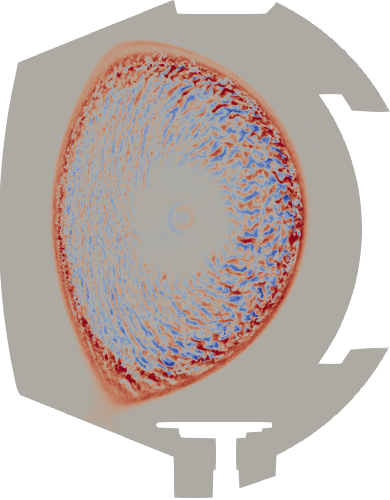}
    \caption{Source Field}
    \end{subfigure} \hfill
    \begin{subfigure}{0.23\linewidth}
    \centering
    \includegraphics[width=\linewidth]{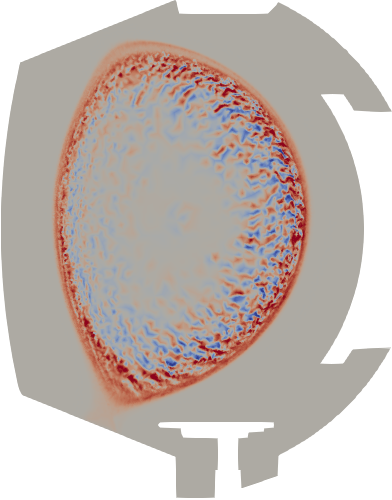}
    \caption{Mesh Intersection}
    \end{subfigure} \hfill
    \begin{subfigure}{0.23\linewidth}
    \centering
    \includegraphics[width=\linewidth]{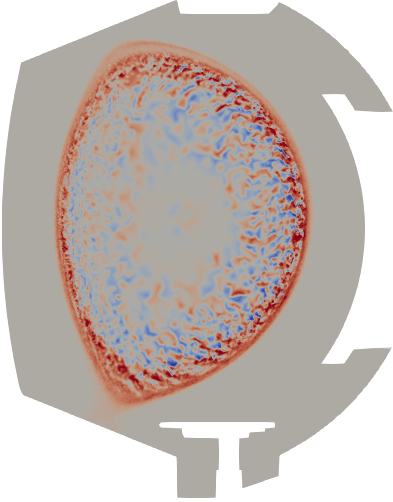}
    \caption{Nearest Neighbor}
    \end{subfigure} \hfill
    \begin{subfigure}{0.23\linewidth}
    \centering
    \includegraphics[width=\linewidth]{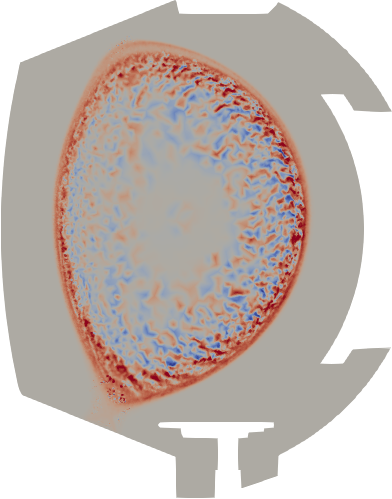}
    \caption{Radial Basis}
    \end{subfigure}
    \caption{Comparison of field transfer of a representative Ion density field from an adiabatic XGCm simulation of the WEST.}
    \label{fig:west-field-transfer}
\end{figure}

To demonstrate the generalizability of these methods to fusion simulation data, we perform a field transfer on a representative ion density field derived from an adiabatic XGCm \cite{zhang2023development} simulation of the WEST reactor (Fig.~\ref{fig:west-field-transfer}a-d). The source and target meshes for this field transfer example are shown in Fig.~\ref{fig:west-meshes}. The source mesh is an XGCm mesh with 611,000 elements that is field aligned in the core region, and only has a depth of a single element between each flux curve. The target mesh is a general unstructured mesh with 120,000 elements. It is clear that each of the methods considered does diffuse the source field in the core region as is expected due to the lower mesh resolution in the target mesh in the core. This eliminates some of the finer scale mode structures. Qualitatively, the target field with the mesh intersection method maintains a better spatial localization than the other methods. For the particular choice of parameters, the field transfer using radial basis functions has some points near the separatrix that look qualitatively incorrect, causing a speckle pattern. This indicates one of the pitfalls of the radial basis function method is that it often requires regularization, and parameter tuning to obtain high-quality field transfers.

To quantitatively evaluate the quality of the field transfers, we show the accuracy and conservation errors for the WEST data shown in Fig.~\ref{fig:west-field-transfer}. On this real data, mesh intersection gives ~20\% improvement over the radial basis function and nearest neighbor methods. However, the real benefit of the mesh intersection method is demonstrated in the conservation error which is seven orders of magnitude better than the radial basis function method and nine orders of magnitude better than the nearest neighbor method.

\begin{table}[]
\centering
\caption{Accuracy and conservation error for ion density field shown in Fig.~\ref{fig:west-field-transfer}.}
\label{tab:west-errors}
\begin{tabular}{|l|l|l|}
\hline
\textbf{Method} & \textbf{Accuracy Error} & \textbf{Conservation Error} \\ \hline
Mesh Intersection              & $4.168\times 10^{-1}$   & $3.985\times 10^{-12}$      \\ \hline
Radial Basis Function             & $5.022\times 10^{-1}$   & $6.738\times 10^{-5}$       \\ \hline
Nearest Neighbor              & $5.114\times 10^{-1}$   & $2.232\times 10^{-3}$       \\ \hline
\end{tabular}
\end{table}

\section{Particle Methods on Unstructured Meshes}\label{Sec:ParticlesMesh}

An important class of multiscale methods employs particle tracking to capture fine-scale behaviors that are coupled to continuum fields discretized over the domain of interest. Effective implementations of these methods must execute at scale on GPU-accelerated computing systems. An increasing number of these particle codes are taking, or plan to take, advantage of unstructured meshes to support continuum-level field calculations. While unstructured meshes offer greater flexibility and accuracy, they also require more complex data structures and algorithms to achieve scalable parallel performance.

To support the development of particle-based, unstructured-mesh analysis codes, the PUMI-Particle infrastructure has been, and continues to be, developed~\cite{diamond2021pumipic}. Key attributes of PUMI-Particle include:
\begin{itemize}
    \item Distributed support for both mesh and particle data across multiple MPI processes.
    \item Scalable execution using two levels of parallelism on GPU-powered HPC systems.
    \item Performance portability achieved by implementing PUMI-Particle on top of the Kokkos library~\cite{edwards2014kokkos}.
    \item Coordination between the mesh and the analysis geometric model using the mesh–model interactions discussed in Section~\ref{Sec:GeometryMeshing}, coupled with efficient adjacency-based search methods~\cite{diamond2021pumipic}.
    \item Multiple particle data structures suited to various forms of particle distributions~\cite{diamond2021pumipic}.
    \item An application programming interface that supports a full range of operations of particles with mesh entities and geometric model entities.
\end{itemize}

A common approach to developing high-performance parallel particle/mesh codes is to employ a distributed particle data structure while maintaining a complete copy of the mesh and continuum fields on each process. This approach is acceptable when the average number of particles per element is large and the required meshes are small enough that storing the entire mesh on each process is feasible. However, when the memory requirements for the mesh and its fields become too large, the mesh itself must be distributed.
Distributing both the particle and mesh data structures substantially increases the complexity of implementing efficient parallel operations on GPU-powered systems. This is because mesh-to-particle and particle-to-mesh operations require increased interprocess communication.

\begin{figure}
\centering
\captionsetup{width=\linewidth, justification=centering}
    \includegraphics[width=0.5\linewidth]{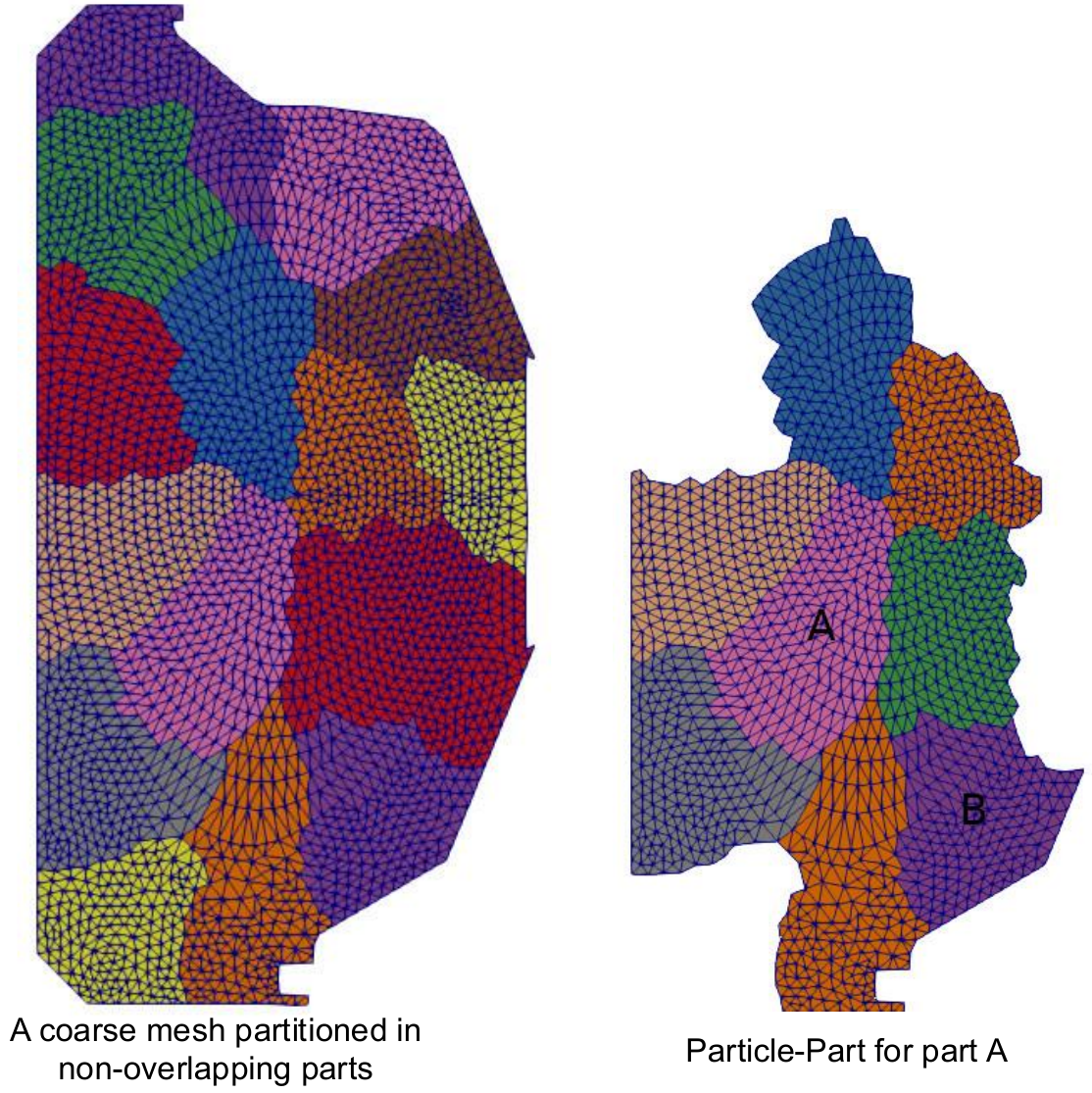}
    \caption{Left image: A coarse mesh partitioned into a set of non-overlapping parts. Right image: The Particle-part for core part A that consist of part A, the parts topologically adjacent to it and any non-adjacent part that is close in terms of topological distance.}
    \label{fig:Particle-Part}
\end{figure}

To reduce the volume of interprocess communication, it is desirable to employ a mesh distribution with sufficiently large buffers of mesh entities. A standard method for defining such buffers is to first partition the mesh into a set of non-overlapping parts, as shown in the left image of Figure~\ref{fig:Particle-Part}, and then add one or more layers of adjacent mesh elements as buffers. Although the distance particles move in a single time step is small, ensuring that most particles remain within elements in a part plus its buffer would require managing buffers consisting of several layers of elements. In this approach, buffer management is performed at the element level, which proves to be inefficient when multiple-element-thick buffers are used.

A more effective alternative is to define a set of \textit{Particle-Parts}, where each Particle-Part consists of a core part together with all topologically adjacent parts, as well as parts that are within a specified number of element layers from the core part. The right image of Figure~\ref{fig:Particle-Part} shows the Particle-Part for part A from the mesh in the left image. The unlabeled parts in this example Particle-Part are adjacent to part A, while part B is also included because it is only three elements away in a topological sense. Although the Particle-Part in this coarse mesh example represents a substantial fraction of the total mesh, in realistic cases with orders of magnitude more elements and parts, each Particle-Part occupies only a small fraction of the total mesh.

Different applications require varying numbers of particles per element and exhibit diverse particle distributions. For example, the XGC edge plasma code~\cite{kuSolve2018} uses several thousand particles per element with a roughly uniform distribution throughout the domain, whereas the GITRm impurity transport code~\cite{nath20233d} exhibits nonuniform and evolving particle distributions, with many elements containing no particles and others having particle counts that change significantly over the course of a simulation. To support this range of applications, PUMI-Particle provides multiple particle data structures organized into two classes. One class associates particles with the elements they currently occupy~\cite{diamond2021pumipic}. This set of data structures is well suited to applications with a large number of particles per element. The second class employs a process-level distribution of particles, with each particle maintaining knowledge of the element it occupies. This set of data structures is better suited to nonuniform distributions where the number of particles per element is small. Within each class, alternative element-based or process-based data structures can be selected to optimize performance for specific applications.

The integration of PUMI-Particle into applications codes is facilitated by an API that supports a set of particle/mesh and mesh/model interactions. Typical particle/mesh API operators include locating the element a particle is in, evaluate forces on particle based on the mesh fields, updating local field values based on the particles properties, determining and counting particle element boundary intersections, etc. The mesh/model interactions, supported by classification and reverse classification (see section~\ref{Sec:GeometryMeshing}), are used to support operations such as finding particles in volume elements that have any mesh face, edge or vertex classified on a model boundary. This capability is useful for determining when a particle is in the scrape off layer near a plasma facing component, where additional physics must be considered in determining the forces acting on the particle.  

\begin{figure}
\centering
\captionsetup{width=\linewidth, justification=centering}
    \includegraphics[width=0.8\linewidth]{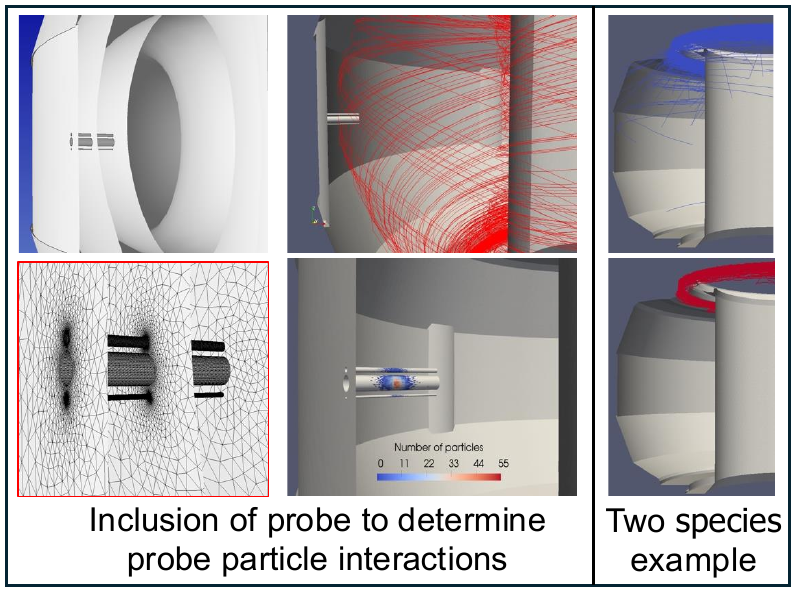}
    \caption{DIII-D example including probe and multiple species of impurities}
    \label{fig:GITRm-result}
\end{figure}

In collaboration with the developer of the GITR impurity transport code~\cite{younkin2021gitr}, PUMI-Particle has been used to develop a fully 3D version, referred to as GITRm~\cite{nath20233d, nath2025gpu}. In a GITRm simulation, the initial set of impurity particles is typically concentrated in elements adjacent to selected plasma-facing components. These particles rapidly redistribute throughout the domain, sometimes reconcentrating in regions of impurity collection. To handle this dynamic behavior, GITRm employs a process-based particle data structure and a dynamic load-balancing procedure that accounts for evolving particle distributions. GITRm also includes an effective multi-species algorithm that maintains a nearly constant total impurity particle count by alternating between modes in which the number of particles increases or decreases at each step.
Figure~\ref{fig:GITRm-result} shows the use of GITRm for modeling particle–probe interactions and tracking multiple impurity species.

\begin{figure}
\centering
\captionsetup{width=\linewidth, justification=centering}
    \includegraphics[width=0.8\linewidth]{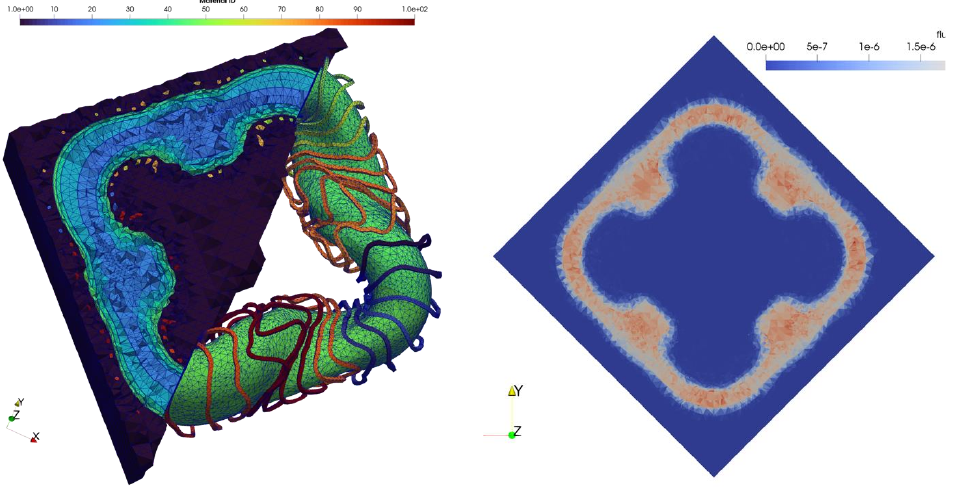}
    \caption{Test case of PUMI-Tally on a ParaStell~\cite{moreno2024parastell} parametrized model of the WISTELL-D stellarator.}
    \label{fig:GITRm-stellarator}
\end{figure}

The design of fusion energy systems includes the application of neutronics simulations. Given that tally calculations are the dominant computational component of many neutronics simulations, a PUMI-Particle-based procedure for executing tally operations on unstructured meshes was developed and integrated into a version of OpenMC~\cite{romano2015openmc, OpenMC}. This resulted in a substantial speedup over the unstructured mesh capabilities previously available in OpenMC~\cite{hasan2025gpu}.
The extensibility of PUMI-Tally to general 3D geometries is demonstrated on test cases on a 3.9-million-element graded mesh for DIII-D with the Helicon RF antenna, shown in Figure~\ref{fig:GITRm-antenna}, and recent simulations of the WISTELL-D stellarator (see Figure~\ref{fig:GITRm-stellarator}).

\section{Closing Remarks}\label{Sec:ClosingRemarks}

Critical to the design of fusion energy systems is the execution of simulation workflows that address complex multiphysics and multiscale behaviors over geometrically complex domains. Given analysis tools capable of addressing each of the required physics behaviors, the development of such workflows must include methods for constructing the inputs to individual analysis codes and for coupling the relevant physics fields. The focus of this paper has been the capabilities required to construct inputs to these analysis codes and to couple them within a modern integrated design process. 

Emphasis has been placed on supporting general geometries by leveraging available computer-aided design and computer-aided engineering tools. Methods for constructing  CAD-based analysis geometric models by combining and manipulating geometry from a variety of sources have been outlined. Such models can be used by today’s computer aided engineering tools to generate the domain discretizations, the meshes, used by the physics analysis codes. Of particular interest are analysis codes that employ unstructured meshes, for which fully automatic mesh generation and adaptive mesh update procedures can be used to automate the execution of simulation workflow steps. 

Since many simulation workflows require the coupling of multiple physics analysis codes, effective code coupling capabilities are essential. In fusion energy system applications this often involves coupling complex physics analysis codes executing on large-scale parallel computing platforms. An approach, together with an associated code-coupling tool, capable of coupling such codes, while ensuring the accurate transfer of solution fields, has been outlined.  

A number of advanced plasma physics analysis codes address the multiscale nature of fusion physics behaviors by combining mesh-based continuum methods with particle methods. When such simulations are executed over complex domains, it is desirable to employ unstructured mesh discretizations. To support such tightly coupled simulations, an API driven tool in which both the mesh and particles can be distribute has been developed and is currently being used in physics studies~\cite{kumar2026understanding}.

\pagebreak
\section*{Acknowledgments}
This research is supported by multiple grants from the U.S. Department of Energy, Office of Science Office of Advanced Scientific Computing Research, Scientific Discovery through the Advanced Computing (SciDAC) Program including the FASTMath SciDAC Institute (DE-SC0021285 \& DE-AC52-0TNA27344), and Fusion Energy Science partner subcontracts including StellFoundry: High-fidelity Digital Models for Fusion Pilot Plant Design (DE-AC02-09CH11466), Computational Evaluation and Design of Actuators for Core-Edge Integration (CEDA) (DE-AC02-09CH11466), HifiStell: High-Fidelity Simulations for Stellarators (DE-SC0024548) and Center for Advanced Simulation of RF, Plasma, Material Interactions (DE-SC0024369).  
This research is also supported through a DOE ASCR SBIR entitled Geometry and Meshing Technologies to Support Fusion Energy System Simulations (DE-SC0024838). 
This work used the resources of the Leadership Computing Facility at the Oak Ridge National Laboratory under DOE Contract No. DE-AC05-00OR22725, and the National Energy Scientific Computing Center (NERSC) at the Lawrence Berkeley National Laboratory (awards FES-ERCAP-m4505, FES-ERCAP-m4564). This research is also supported through a DOE Fusion Innovation Research Engine (FIRE) Collaboratives Program through a grant titled Mitigating Risks from Abrupt Confinement Loss (MIRACL) (DE-AC02-09CH11466).
Any opinions, findings, and conclusions or recommendations expressed are those of the author(s) 
and do not necessarily reflect the views of the U.S. Department of Energy. We gratefully acknowledge use of the research computing resources of the Empire AI Consortium, Inc, with support from the State of New York, the Simons Foundation, and the Secunda Family Foundation.

\pagebreak
\bibliographystyle{style/ans_js}                                                                           
\bibliography{bibliography}

\end{document}